\documentclass[sn-aps,iicol]{sn-jnl}

\jyear{2023}%

\theoremstyle{thmstyleone}%
\theoremstyle{thmstyletwo}%

\theoremstyle{thmstylethree}%

\usepackage{amsmath,amssymb}
\usepackage{graphicx}
\usepackage{dcolumn}
\usepackage{bm}
\usepackage{verbatim}
\usepackage{tensor}
\usepackage{bbm}
\usepackage{tabularx}
\usepackage{slashed}
\usepackage{cancel}

\def\be{\begin{equation}}
\def\ee{\end{equation}}
\def\bea{\begin{eqnarray}}
\def\eea{\end{eqnarray}}
\def\ba#1\ea{\begin{align}#1\end{align}}
\def\bg#1\eg{\begin{gather}#1\end{gather}}
\def\bm#1\em{\begin{multline}#1\end{multline}}
\def\bmd#1\emd{\begin{multlined}#1\end{multlined}}

\begin{document}

\title[\,]{Revisiting the electronic properties of disclinated graphene sheets}


\author*[1]{\fnm{Nicol\'as } \sur{Fernández}}\email{nicolas.fernandez@unmsm.edu.pe}

\author[2]{\fnm{Pierre} \sur{ Pujol}}\email{pierre.pujol@irsamc.ups-tlse.frmar}

\author[3,4]{\fnm{Mario} \sur{Solís}}\email{mario.solis@ib.edu.ar}

\author[5]{\fnm{Teófilo} \sur{Vargas}}\email{tvargasa@unmsm.edu.pe}

\affil[1]{\orgdiv{Facultad de Ciencias F\'isicas}, \orgname{Universidad Nacional Mayor de San Marcos}, \orgaddress{\street{Avenida Venezuela s/n Cercado de Lima}, \city{Lima}, \postcode{ 15081}, \state{Lima}, \country{Perú}}}

\affil[2]{\orgdiv{Laboratoire de Physique Théorique}, \orgname{CNRS and Université de Toulouse, UPS}, \\
\orgaddress{ \city{Toulouse}, \postcode{F-31062},  \country{France}}}

\affil[3]{\orgdiv{Centro Atómico Bariloche}, \orgname{CNEA and CONICET}, \orgaddress{ \city{S. C. de Bariloche}, \\
\postcode{R8402AGP}, \state{Rio Negro}, \country{Argentina}}}

\affil[4]{\orgdiv{Instituto Balseiro}, \orgname{UNCuyo}, \orgaddress{ \city{S. C. de Bariloche}, \postcode{R8402AGP}, \state{Rio Negro}, \country{Argentina}}}

\affil[5]{\orgdiv{Grupo de F\'isica Te\'orica GFT and Grupo de Astronom\'ia SPACE}, \orgname{Universidad Nacional Mayor de San Marcos}, \orgaddress{\street{Avenida Venezuela s/n Cercado de Lima}, \city{Lima}, \postcode{ 15081}, \state{Lima}, \country{Perú}}}


\abstract{The interplay between topological defects, such as dislocations or disclinations, and the electronic degrees of freedom in graphene has been extensively studied. In the literature, for the study of this kind of problems, it is in general used either a gauge theory or a curved spatial Riemannian geometry approach, where, in the geometric case, the information about the defects is contained in the metric and the spin-connection. However, these topological defects can also be associated to a Riemann-Cartan geometry where curvature and torsion plays an important role. In this article we study the interplay between a wedge disclination in a planar graphene sheet and the properties of its electronic degrees of freedom. Our approach relies in its relation with elasticity theory through the so called elastic-gauge, where their typical coefficients, as for example the Poisson's ratio, appear directly in the metric, and consequently also in the electronic spectrum. 	}




\maketitle


\section{Introduction}\label{sec:intro}
Graphene has received much attention in recent years due to its elastic and electronic properties, which offer key industrial impact in nanoelectronic devices \cite{neto2009electronic}. Since its electronic band structure was studied by Wallace and Semenoff \cite{wallace1947band,semenoff1984condensed}, and then produced by Novoselov et. al. \cite{novoselov2004electric}, a very wide variety of theoretical and experimental work has been conducted. Among them, one can mention the study of the electronic properties and how the electrons interact with, for example, an electromagnetic field, topological defects, disorder or impurities. The most important feature of graphene is the linear energy spectrum at two Dirac points in the Brillouin zone, which provide a low energy description with massless Dirac fermions. Another interesting result comes from the study of material irregularities, which have dramatic effects on its electronic properties.
The topological defects give rise to interesting modifications of the electronic properties. They are classified into two general types: disclinations and dislocations \cite{read1950dislocation,harris1977disclinations}. From the atomistic point of view,  a disclination adds or removes one atom in the lattice, and a dislocation can be viewed as a pair of disclinations. If one considers the system as a continuum medium, it can be  studied using elasticity theory \cite{landau1986theory}. In this approach, one can add or remove a piece of material and then glue the new boundaries. This procedure is known as Volterra's process \cite{puntigam1997volterra}. In both cases, the key quantities that specify these irregularities are the Burgers and Franck vectors: they express the deviation with respect to a defect-free configuration. The presence of these defects introduces an extra field which interacts and modifies the free electronic dynamics. This new field enter in the electron physics either through a gauge field or through a geometric approach.
In the first case, the gauge field describes the modifications of the electronic degrees of freedom due to the presence of disclinations or dislocations, and, in the continuum limit, it acts as an point flux producing an Aharonov-Bohm effect \cite{kawamura1978scattering}. This approach was developed extensively in the last three decades in the context of fullerenes and graphite. The insightful studies of Gonzalez et. al. \cite{gonzalez1992continuum,gonzalez1993electronic} demostrated that the folded carbon lattice mimics a gauge field. An interesting application to graphite system was developed by Lammert and Crespi \cite{lammert2000topological}, where they studied graphite in the presence of a disclination. They viewed this lattice as a graphite cone where the disclination is at its apex. Within this approach they calculated the LDOS (local density of states); similar results were previously reported \cite{tamura1994disclinations} within a tight-binding formulation (see \cite{vozmediano2010gauge} and references therein for a summary of these techniques and applications), and also in the presence of a magnetic field.
In the second case, the topological defects were considered sources of geometric distortions, and these distortions produce a non-trivial geometry background in which the fermionic dynamics develop: the euclidean flat metric is promoted to a curved one. The presence of defects (dislocations or disclinations) produces a singularity point in the manifold which is phenomenologically characterized by a conical metric (see for i.e. \cite{tod1994conical}). This metric can also be related to a cosmic string \cite{vilenkin1981gravitational} in high-energy physics. The conical metric provides the fermions with a spin connection, which plays the role of the gauge field in the previously mentioned approach. In the case of graphene, the application of this technique was reported by Osipov and Kochetov \cite{osipov2001dirac}, where they obtained the solution of the massless Dirac equation in the conical geometry and its LDOS. They reported that there are no extended states with non-zero density of states, and their results showed a different power energy dependence with respect to \cite{lammert2000topological}. Similar studies were done by Vozmediano et. al. \cite{vozmediano2008gauge} where the authors computed the LDOS, or Cortijo et. al. \cite{cortijo2007effects} who solved the Dirac equation for ripples in graphene. However, much of this studies were focused in the interplay between the fermionic dynamics and the geometry instead of the interplay between electronic-elastic properties, or how the previous results in elasticity theory \cite{landau1986theory} are connected in this approach.
An important progress in geometric-elastic properties was achieved by Katanaev and Volovich \cite{katanaev1992theory} (see \cite{katanaev2005geometric} for a review). They relates the Burgers and Franck vectors with torsion and curvature, respectively. The inclusion of torsion was previously mentioned long time ago \cite{kondo1952geometrical,bilby1955continuous,kroner1980continuum}. Disclinations carry curvature but not torsion, while  dislocations carry torsion but not curvature. In their approach, the Riemann-Cartan geometry is used, where the curvature and torsion are non-trivial. The final ingredient related to the elastic properties was finally realized in \cite{katanaev2003wedge}, where the authors implemented an elastic-gauge to derive the elastic wedge disclinated metric from a conical metric \cite{landau1986theory}. The importance of this gauge lies in its explicit dependence on the elastic constants. With this ideas, the interplay between fermions (in graphene) was studied by Furtado et. al. \cite{furtado1994landau,furtado1994binding}. They also included the interaction with an external magnetic field and determined the electronic spectrum. However, the consequences of the elastic-gauge in graphene were not previously reported. As we show below, the elastic-gauge is responsible of an important change in the electronic behavior.
In this work, we study the electronic dynamics within the Katanaev-Volovich approach \cite{katanaev2005geometric} of a planar graphene sheet in the presence of a wedge disclination. We will focus on how its electronic dynamics is changed by the presence of a disclination. Our major goal is the interesting interplay between the electronic and elastic properties that appear in the spectrum. We also compute the LDOS and we find a different power dependence than the one obtained in previous works which focused on a conical shape \cite{osipov2001dirac, lammert2000topological}.
\section{Wedge disclination and its associated metric}\label{sec:conical_def}
In the geometric approach of defects \cite{katanaev2005geometric}, the flat Minkowskian space-time metric is replaced by a curved metric in a Riemann-Cartan manifold; the line element is
\be\label{eq.general_metric}
ds^2= g_{\mu\nu} dx^\mu dx^\nu = dt^2 - g_{ij}dx^idx^j\,,
\ee
where $\mu,\nu$, are the space-time indices and $g_{ij}$ is the spatial curved metric in euclidean signature $(+,+)$. The temporal component of the metric is flat indicating that the topological defects are not time-dependent. In this perspective, the system is characterized by the curvature and torsion tensors
\bea
R_{\mu\nu}\,^{ab} &=& \partial_\mu \omega_\nu\,^{ab} -  \partial_\nu \omega_\mu\,^{ab} +\omega_\mu\,^a_c\omega_\nu\,^{cb} -\omega_\nu\,^a_c\omega_\mu\,^{cb}  \,, \nonumber \\
T_{\mu\nu}\,^a &=& \partial_\mu h_\nu\,^a - \partial_\nu h_\mu\,^a +\omega_\mu\,^a_b h_\nu\,^b  - \omega_\nu\,^a_b h_\mu\,^b\,, 
\eea
where $a,b$ are tangent-space indices, $\omega_\mu\,^{ab}$ is the spin-connection and $h_\mu\,^b$ is an associated tetrad. According to the geometric approach \cite{katanaev1992theory,katanaev2003wedge,katanaev2005geometric}, the Burgers and Franck vectors are associated to 
\bea
b^a =\int_S dx^\mu \wedge dx^\nu\, T_{\mu\nu}\,^a \,, \\
\Omega^{ab} =\int_S dx^\mu \wedge dx^\nu \, R_{\mu\nu}\,^{ab}\,,
\eea
in this context, the presence of a non-trivial torsion indicates the presence of dislocation while the curvature is associated to disclinations.
The Action for topological defects was obtained in \cite{katanaev1992theory, katanaev2005geometric}, where in the absence of disclinations, the curvature vanish. In this case, the Riemann-Cartan geometry is reduced to a space of absolute parallelism giving an analogue of teleparallel gravity. In this theory, the Riemann curvature is determined by the torsion tensor which depends on the Weitzenb\"ock connection (see \cite{arcos2004torsion,bahamonde2022teleparallel} and references therein for more details). In the rest of the paper we use the notation of \cite{arcos2004torsion}, where $\overset{\circ}{}$, $\overset{\bullet}{}$ denote the quantities given by the Levi-Civita and Weitzenb\"ock connection, respectively. 
The equation of motion for geometry in the presence of a disclination is
\be\label{eq.Einstein_Eq}
\overset{\circ}R_{\mu\nu}- \frac{1}{2} g_{\mu\nu}\overset{\circ}R = \kappa \Theta_{\mu\nu}\,,
\ee
where $\kappa$ is a constant which makes the equation well defined from a dimensional point of view \footnote{In gravity context is $\kappa = 8\pi G_N$.} and $\Theta_{\mu\nu}$ is the energy-momentum tensor. The only non-vanishing component of the energy-momentum tensor for a disclination at the origin is
\be\label{eq.energy_momentum_tensor}
\Theta_{zz}= \frac{2\pi}{\sqrt{g}} \theta \delta^2(x)\,,
\ee
Which depend explicitly on the deficit angle $\theta$. The Dirac delta in eq. (\ref{eq.energy_momentum_tensor}) produce a singularity in the geometry. Following the same steps of \cite{katanaev1992theory, katanaev2005geometric},  the metric, written in cylindrical coordinates, is given by
\be\label{eq.conical_metric_1}
g_{ij}dx^idx^j = d\tilde{r}^2 + \alpha^2\tilde{r}^2d\varphi^2 \,,
\ee
where $0 \leq \tilde{r} <\infty$ and $0 \leq \varphi <2\pi$.The parameter $\alpha$ is related with the deficit angle $\theta$ of disclination as $\alpha = 1 +\theta$.
One way to see how the metric in eq. (\ref{eq.conical_metric_1}) arise because of the wedge disclination is to consider the Volterra's process \cite{puntigam1997volterra}. First, consider the system as a disk, then remove a wedge piece of the material with angle $\theta$. After that, glue the two sides while maintaining the system flat, and the disclination core is located at the center of the disk. An important point that deserves a clarification is that, in previous works, the metric used for a system with conical geometry do not contain information about the elasticity parameters \cite{landau1986theory}. This gave rise to the proposal of Katanaev \cite{katanaev2003wedge} to use what he called an a elastic-gauge, in which the elasticity parameters appear explicitly in the metric. For the planar geometry we consider here we follow this route and use the metric that contains the elasticity parameters. With this choice, we find
\be\label{eq.r_to_Katanaev}
\tilde{r}= \frac{r^{\gamma}}{\tilde\gamma R^{\gamma-1}} \equiv \rho\,, 
\ee
where $R$ is the radii of the disk (or cylinder base) and
\be\label{eq.gamma_tilde}
\gamma=  \frac{-\theta \sigma +\sqrt{\theta^2 \sigma^2 + 4(1 + \theta)(1 - \sigma)}}{2(1 - \sigma)}\,,
\ee
here the Poisson's ratio is denoted by $\sigma$. As a result, the elastic-gauge relates the general conical metric (\ref{eq.conical_metric_1})  with a specific conical metric containing information of the elastic properties of the sample and reproduce a well known result \cite{landau1986theory} in elasticity theory. Hence, using eq. (\ref{eq.r_to_Katanaev}) in (\ref{eq.conical_metric_1}) , we find
\be\label{eq.katanaev_metric}
 g_{ij}dx^idx^j = d\rho^2 + \alpha^2\rho^2d\varphi^2\,,
\ee
with $0 \leq r < R$, $0 \leq \varphi < 2 \pi$, $\alpha = 1 + \theta$, where we used the variable $\rho$ defined in (\ref{eq.r_to_Katanaev}).
Therefore, the complete metric (\ref{eq.general_metric}) is
\be\label{eq.cone_metric_2}
ds^2 = dt^2 - d\rho^2 - \alpha^2\rho^2d\varphi^2\,,
\ee
where $0< \rho \leq R/\gamma$, $0<\varphi <2\pi$, $\alpha = 1 + \theta$ and $\rho\,, \gamma$ are given by eqs. (\ref{eq.r_to_Katanaev}) and (\ref{eq.gamma_tilde}) respectively. This result is extremely important for the next sections, as it modifies the electronic spectrum and LDOS as we show below.
\section{Wedge disclinations and Dirac equation}\label{sec:Dirac}
The metric in eq. (\ref{eq.cone_metric_2}) provides a natural description of the considered defect. We now can turn to the study of the electronic degrees of freedom in an elastic medium with a wedge disclination from the perspective of the Weitzenb\"ock geometry. In this perspective we introduce the Dirac equation in the farmework of the teleparallel Fock-Ivanenko covariant derivative \cite{arcos2004torsion,bahamonde2022teleparallel} . The equation we want to study is:
\begin{equation}\label{eq.Dirac_eq_1}
    i \gamma^\mu \overset{\bullet}{\mathcal{D}}\tensor{}{_\mu} \psi - m\psi = 0\,,
\end{equation}
where we set $\hslash = v_F = 1$. Here, 
\begin{equation}
    \overset{\bullet}{\mathcal{D}}\tensor{}{_\mu} = \partial_\mu + \frac{i}{2} \overset{\bullet}{K}\tensor{}{_\mu^a^b} \tensor{S}{_a_b}\,,
\end{equation}
where $\overset{\bullet}{K}\tensor{}{_\mu^a^b}$ is the contorsion tensor, $\tensor{S}{_a_b} = \frac{i}{4} \left[ \gamma_a , \gamma_b \right]$ is the Lorentz spin-1/2 generator, and $\gamma^\mu = \tensor{h}{^\mu_a} \gamma^a$ are the curved Dirac matrices in terms of a nontrivial tetrad field $\tensor{h}{^\mu_a}$. The spacetime and the tangent-space metrics are related by $\tensor{g}{_\mu_\nu} = \tensor{\eta}{_a_b} \tensor{h}{^a_\mu} \tensor{h}{^b_\nu}$ and the contorsion tensor can be determined by the following relation
\begin{equation}
   \overset{\bullet}{K}\tensor{}{_\mu^a^b} = \tensor{h}{^a_\rho} \overset{\bullet}{K}\tensor{}{^\rho_\mu_\nu} \tensor{h}{_c^\nu} \tensor{\eta}{^c^b}\,,
\end{equation}
with $\overset{\bullet}{K}\tensor{}{^\rho_\mu_\nu}=\overset{\bullet}{\Gamma}\tensor{}{^\rho_\mu_\nu} -\overset{\circ}{\Gamma}\tensor{}{^\rho_\mu_\nu}$, where $\overset{\bullet}{\Gamma}\tensor{}{^\rho_\mu_\nu}=\tensor{h}{_a^\rho} \partial_\mu \tensor{h}{^a_\nu}$ is the Weitzenb\"ock connection and $\overset{\circ}{\Gamma}\tensor{}{^\rho_\mu_\nu} = \frac{1}{2} \tensor{g}{^\sigma^\rho} \left( \partial_\mu \tensor{g}{_\rho_\nu} + \partial_\nu \tensor{g}{_\rho_\mu}-\partial_\rho \tensor{g}{_\mu_\nu} \right)$ is the Levi-Civita connection of the Riemannian geometry. 
The Dirac equation (\ref{eq.Dirac_eq_1}) can be written in the Schr\"odinger's form as
\begin{equation}
i \tensor{\partial}{_t} \psi = \left( \tensor{\gamma}{^t} m - i \tensor{\gamma}{^t} \gamma^i\partial_i - i \tensor{\gamma}{^t} \gamma^i \overset{\bullet}{K}\tensor{}{^a^b_i} \ \tensor{S}{_a_b} \right)\psi\,.
\end{equation}
The Dirac matrices obey the Clifford's algebra
\begin{equation}
\left\{ \gamma^\mu, \gamma^\nu \right\} = 2 g^{\mu\nu}\,,
\end{equation}
and can be chosen to be the Pauli matrices $ \gamma^t = \gamma^0 = \sigma^3$, $\gamma^1 = i \sigma^2$ and $\gamma^2 = - i \sigma^1$, or explicitly
\be\label{eq.Dirac_Matrices}
\gamma^t = \left(\begin{matrix}
    1 & 0  \\
    0 & -1 \\
\end{matrix} \right)\,, \quad
\gamma^1 = \left( \begin{matrix}
    0 & 1  \\
    -1 & 0 \\
\end{matrix} \right)\,, \quad
\gamma^2 = \left( \begin{matrix}
    0 & -i  \\
    -i & 0 \\
\end{matrix}\right)\,,
\ee
with this choice we find: $\overset{\bullet}{K}\tensor{}{^1^2_\varphi} = 1 - \alpha$ and $\tensor{S}{_1_2} =\frac{1}{2} \sigma^3$. Therefore, the Dirac Hamiltonian is given by
\begin{equation}
    H = \tensor{\gamma}{^t} m - i \tensor{\gamma}{^t} \Gamma (\varphi) \partial_\rho - i \tensor{\gamma}{^t} \Gamma' (\varphi) \frac{\partial_\varphi}{\alpha \rho} + \frac{1-\alpha}{2 \alpha \rho} \tensor{\gamma}{^t} \Gamma' (\varphi) \sigma^3\,,
\end{equation}
where $\Gamma (\varphi) = \gamma^1 \ cos(\varphi) + \gamma^2 \ sin(\varphi)$ and $\Gamma' (\varphi) = - \gamma^1 \ sin(\varphi) + \gamma^2 \ cos(\varphi)$, as shorthand notation.
Inspired in the result of \cite{furtado2008geometric}, in the process of a parallel transport of a spinor along a closed trajectory C around the defect, the holonomy matrix $U(C) = \text{exp} \left( - \oint_C \Gamma_\mu (x) dx^\mu \right)$, being $\Gamma_\mu$ is the spinorial connection, provide the quantum phase acquired by the wave function, i.e., $\psi = U(C) \psi_0$, and the Hamiltonian becomes
\begin{equation}\label{Dirac-Hamiltonian}
    H = \tensor{\gamma}{^t} m - i \tensor{\gamma}{^t} \Gamma (\varphi) \partial_\rho - i \tensor{\gamma}{^t} \Gamma' (\varphi) \frac{\partial_\varphi}{\alpha \rho}\,.
\end{equation}
Noticing that the total angular momentum $J = -i \partial_\varphi + \frac{1}{2} \sigma^3$ commutes with the Hamiltonian (\ref{Dirac-Hamiltonian}), the eigenfunctions are classified by the eigenvalues $j = l \pm 1/2$, $l = 0, \pm 1, \pm2, \dots$. We can therefore analyze the stationary state of the Dirac equation using the following anzatz
\begin{equation}
\psi_0 = e^{-i E t} \left(\begin{array}{c}
u(\rho) e^{i l \varphi} \\
\text{v}(\rho) e^{i(l+1) \varphi}
\end{array}\right)\,,
\end{equation}
which give the following system of differential equations
\bea
(E-m) u(\rho) &=& - i \partial_\rho \text{v}(\rho) - i \frac{l+1}{\alpha \rho} \text{v}(\rho)\,, \\
(E+m) \text{v}(\rho) &=& - i \partial_\rho u(\rho) + i \frac{l}{\alpha \rho} u(\rho)\,.
\eea
In the following, we focus on the case of massless chiral fermion in order to describes the low energy electronic degrees of freedom of graphene.
\section{Massless fermions in disclinated planar Graphene}\label{sec:WedgeGraphene}

We now specify our previous system of differential equations on massless fermions, which is the relevant case for the description of graphene. Solving for $u(\rho)$ we get
\begin{equation}\label{u-function}
    u(\rho) = C_1 \rho^{\xi} J_\eta (E \rho)\,,
\end{equation}
where we consider only the Bessel function of the first kind as it is the only one finite at the origin. In this section we use $\xi = \frac{\alpha - 1}{2 \alpha}$ and $\eta = \pm \frac{ \alpha - 2 l -1 }{2 \alpha}$. Then $\text{v}(\rho)$ is given by
\begin{equation}\label{v-function}
    \text{v}(\rho) = \pm i C_1 \rho^{\xi} J_{\tilde{\eta}} (E \rho)\,,
\end{equation}
where $C_1$ is a constant to be determined below, $\tilde{\eta}= \pm (\eta + 1)$ and the signs $\pm$ corresponds to positive and negative energy solutions, respectively. Hence, there are two independent solutions for each energy state, one for $\eta, \tilde{\eta}>0$ and another for $\eta, \tilde{\eta}<0$.
Now we focus in case of positive energy and we want to obtain the constant $C_1$. The normalization condition is written as
\begin{equation}
\int \left( \lvert u(\rho) \rvert^2 + \lvert \text{v}(\rho) \rvert^2 \right) \sqrt{-g} \ d \rho \ d \varphi = 1\,,
\end{equation}
where $g$ is the determinant of the metric. Using the equations (\ref{u-function}) and (\ref{v-function}) find
\begin{equation} \label{normalization}
2\pi \alpha {C_1}^2 \int_0^{R/\gamma} \rho^{2 \xi + 1} \left( {J_\eta}^2 (E \rho) + {J_{\eta + 1}}^2 (E \rho) \right) d \rho = 1\,,
\end{equation}
and we obtain 
\begin{equation}
  C_1 = \frac{1}{E^\eta}\sqrt{f(\eta, \xi, R, \gamma) \ {}_2 F_3\left( \zeta \right)}\,,
\end{equation}
where $f(\eta, \xi, R, \gamma)$ is a constant whose dependence on the parameters do not need to be made explicit here. The ${}_2 F_3\left( \zeta \right)$ is an Hyper-geometric function of arguments $\zeta = \{ \eta + 1 + \xi, \eta + 1/2; 2\eta + 1, \eta + 1, 2 + \eta + \xi; - (R/\gamma)^2 E^2 \}$. For fixed values of $E$ and taking $R \rightarrow \infty$ we get
\begin{equation}\label{C1constant}
  C_1 \propto E^{1/2} \left( \frac{\gamma}{R} \right)^{\xi+\frac{1}{2}}
\end{equation}
Reanalyzing the normalization condition (\ref{normalization}) for $E \rho \ll 1$, we find 
\begin{equation}
\int_0^{R/\gamma} \rho^{2 \xi + 1} \left( \left( \frac{E \rho}{2} \right)^{2\eta} + \left( \frac{E \rho}{2} \right)^{2 \tilde{\eta}} \right) d \rho = 1\,,
\end{equation}
The integrability of the previous expression induce restrictions in the values of the angular momentum quantum number as: $l< \text{min}(2 \theta +1, 3\theta +2)$ for $\eta, \tilde{\eta} > 0$ and $l > \text{max}(-1-\theta, 0)$ for $\eta, \tilde{\eta} < 0$. Furthermore, the parameter $\alpha$ is bounded by $0< \alpha <1$, because of the removal of a part of the material.
To study the wave function near the center of the disk, it is useful to consider small arguments in (\ref{u-function}), (\ref{v-function}), i.e., with fixed values of $E$ and taking $\rho \rightarrow 0$. For $\eta, \tilde{\eta} >0$, the dependence on $E$ and $\rho$ is
\begin{equation}
\left(\begin{array}{l}
u(\rho) \\
\text{v}(\rho)
\end{array}\right) \sim \left(\begin{array}{l}
E^{1/2+\eta} \rho^{\xi+\eta} \\
E^{1/2+\tilde{\eta}} \rho^{\xi+\tilde{\eta}}
\end{array}\right)\,,
\end{equation}
and for $\eta, \tilde{\eta} < 0$,
\begin{equation}
\left(\begin{array}{l}
u(\rho) \\
\text{v}(\rho)
\end{array}\right) \sim \left(\begin{array}{l}
E^{1/2-\eta} \rho^{\xi-\eta} \\
E^{1/2-\tilde{\eta}} \rho^{\xi-\tilde{\eta}}
\end{array}\right)\,.
\end{equation}
We can express the above results in terms of the deficit angle $\theta$, we obtain 
\begin{equation}
\left(\begin{array}{c}
u(\rho) \\
\text{v}(\rho)
\end{array}\right) \sim \left(\begin{array}{l}
E^{\frac{1+2\theta-2 l}{2+2\theta}} \rho^{\frac{\theta - l}{1 + \theta}} \\
E^{\frac{3+4\theta-2 l}{2+2\theta}} \rho^{\frac{2\theta - l + 1}{1 + \theta}}
\end{array}\right), \quad \eta, \tilde{\eta} > 0\,,
\end{equation}
and
\begin{equation}
\left(\begin{array}{c}
u(\rho) \\
\text{v}(\rho)
\end{array}\right) \sim \left(\begin{array}{l}
E^{\frac{1+2 l}{2+2\theta}} \rho^{\frac{l}{1 + \theta}} \\
E^{\frac{-2\theta+2 l-1}{2+2\theta}} \rho^{\frac{-\theta + l - 1}{1 + \theta}}
\end{array}\right), \quad \eta, \tilde{\eta} < 0\,.
\end{equation}
In particular, with the meaningful values for the angle $\theta = 0, -1/6, -1/3$, we get $l = 0$ and, therefore, $\psi \sim E^{1/2}$, $\psi \sim E^{2/5} \rho^{-1/5}$, and $\psi \sim E^{1/4} \rho^{-1/2}$, respectively. In the same way, for $\theta = 1/6, 1/3, 1/2$, we get $l = 1$ and $\psi \sim E^{2/7} \rho^{-1/7}$, $\psi \sim E^{1/8} \rho^{-1/4}$ and $\psi \sim \rho^{-1/3}$, respectively.

In order to compute the local density of states, we notice that it diverges for $\rho \to 0$. This divergence can be regularized by considering the total density of states on a small disk $0< \rho \leq \delta$, as was done in \cite{osipov2001dirac} and \cite{lammert2000topological}. The local density of states is defined as
\begin{equation}
\rho(\mathbf{r}, E) = \sum_i \lvert \psi_i (\mathbf{r}) \rvert^2 \delta(E - E_i)\,,
\end{equation} 
thus, we calculate the total density of states as
\begin{equation}
DoS = \left( \int_0^{\frac{\delta^\gamma}{\gamma R^{\gamma -1}}} u(\rho) d\rho \right)^2 + \left( \int_0^{\frac{\delta^\gamma}{\gamma R^{\gamma -1}}} v(\rho) d\rho \right)^2 \,,
\end{equation}
and we find the dependence for $E$ and $\delta$ as
\begin{equation}\label{eq.DoS_1}
DoS \propto \left\{ \begin{array}{ll}
E^{\frac{4\theta -2 l +3}{1+\theta}} \delta^{\gamma \frac{6\theta-2 l +4}{1 + \theta}} & \quad \eta, \tilde{\eta} > 0\,, \\[10pt]
E^{\frac{-2\theta +2l-1}{1 + \theta}} \delta^{\gamma \frac{2l}{1 + \theta}} & \quad \eta, \tilde{\eta} < 0\,.
\end{array} \right.
\end{equation}
In the leading order for equation (\ref{eq.DoS_1}), we obtain
\begin{equation}\label{eq.LDOS_values}
DoS \propto \left\{ \begin{array}{llllll}
E^{1/2}\delta^{\gamma}\,, & \quad \theta = -1/3, \quad l = 0\,, \\[10pt]
E^{4/5}\delta^{8\gamma/5}\,, & \quad \theta = -1/6, \quad l = 0\,, \\[10pt]
E \delta^{2\gamma}\,, & \quad \theta = 0, \quad l = 0\,, \\[10pt]
E^{4/7}\delta^{12\gamma/7}\,, & \quad \theta = 1/6, \quad l = 1\,, \\[10pt]
E^{1/4} \delta^{3\gamma/2}\,, & \quad \theta = 1/3, \quad l = 1\,, \\[10pt]
\delta^{4\gamma/3}\,, & \quad \theta = 1/2, \quad l = 1\,, \\[10pt]
\end{array} \right.
\end{equation}
for values $\theta = 0, -1/6, -1/3$ of the deficit angle. In the inextensional limit ($\sigma = 1/2$, i.e., rigid membrane), our results are in agreement with previous works, \cite{osipov2001dirac} and \cite{lammert2000topological}.
At last, we study the discretization of spectrum because of the finite size system. We consider the billiard boundary condition \cite{berry1987neutrino} with an infinite potential at the boundary $\rho = R/\gamma$ assuming a circular contour, i.e., with $\varphi = \alpha$ at the boundary. Such a condition for the spinor components can be written as
\begin{equation}\label{boundary-condition}
\frac{\psi_2}{\psi_1} = i e^{i\varphi}\,,
\end{equation}
where $\psi_1 = C_1 \rho^\xi J_\eta (E R/\gamma) e^{i l \varphi} $ and $\psi_2 = i C_1 \rho^\xi J_{\eta + 1} (E R/\gamma) e^{i (l + 1) \varphi}$, so (\ref{boundary-condition}) becomes
\begin{equation}
J_\eta (E R/\gamma) = J_{\eta + 1} (E R/\gamma)\,,
\end{equation}
Considering the asymptotic behavior of the Bessel functions for $E R/\gamma \gg 1$,
\begin{equation}
J_\eta (E R/\gamma) \approx \sqrt{\frac{2}{\pi E R/\gamma}} \cos \left( E \rho - \frac{\eta \pi}{2} -\frac{\pi}{4}\right)\,,
\end{equation}
The boundary condition impose
\begin{equation}
\cos \left( E R/\gamma - \frac{\eta \pi}{2} -\frac{\pi}{4}\right) = \sin \left( E R/\gamma - \frac{\eta \pi}{2} -\frac{\pi}{4}\right)\,,
\end{equation}
hence the argument must be equal to $\pi/4 + n$. This give us the spectrum in the presence of wedge disclination as
\begin{equation}\label{Enery-levels}
E = \frac{\pi \gamma}{R} \left( n + \frac{1}{2}  + \bigg\rvert \frac{\alpha -2l -1}{4 \alpha} \bigg\rvert \right)\,.
\end{equation} 
It is important to point out that the result depends on the elasticity properties through the definition of $\gamma$, given in equation (\ref{eq.gamma_tilde}), with an explicit dependence on the Poisson's ratio.
This last result is to put in perspective with the remark that when a wedge of material is removed, the electrons spread over a smaller area than the one available in a material without defects, whereas, when a wedge of material is added, the electrons spread over a larger area. Geometrically, these correspond to positive and negative curvature surfaces respectively.
%
\section{Conclusions}\label{sec:Conclu}

In this work we showed how  the fermionic dynamics in graphene is affected by a wedge disclination. We used the geometric approach developed in \cite{katanaev2005geometric} and implemented the line element of a wedge disclination found in elasticity theory \cite{landau1986theory, katanaev2005geometric} within the elastic-gauge \cite{katanaev2003wedge}. This choice of metric corresponds to the case where the graphene sheet is kept planar despite the fact of a presence of a disclination. Our main results are the LDOS in equation (\ref{eq.LDOS_values}), where we found a different power dependence in radius $\delta$ than in previous works made on a conical geometry \cite{lammert2000topological,osipov2001dirac} and the finite size spectrum in equation (\ref{Enery-levels}) was obtained using the billiard condition. A similar result was reported in \cite{furtado1994landau} where the authors studied the Landau's levels. However, it is important to stress that in this work, our results contain the explicit presence of the Poisson's ratio, which appear for example in eq. (\ref{Enery-levels}), a result directly arising from the planar geometry which manifest itself through the so called elastic-gauge. As a consequence, the result for the spectrum, eq. (\ref{Enery-levels}) show the explicitly interplay between the electronic and elastic properties in a disclinated graphene sheet. Our results may be verified in an experimental set-up and provide, in principle, a nondestructive manner to compute the Poisson's ratio. We are confident that this analysis can be extended to a graphene sheet with several disclinations with equal or different deficit angles, and the study of the Landau levels of graphene with disclinations. 

\section*{Acknowledgments}
The authors would like the thank the LA CoNGA Physics program which was at the origin of this collaboration. M. S. is supported by CNEA, Conicet and UNCuyo-Instituto Balseiro. T. V. is supported by the Vice-rectorate for Research and Postgraduate B20131101 and CONCYTEC through the Research Academic grant. M. S. would like to thank the Facultad de Ciencias F\'isicas, Universidad Nacional Mayor de San Marcos for its hospitality during the first stage of this project.

\section*{Declarations}

\subsection*{Author contribution}
All authors contributed to the study equally.

\subsection*{Data Availability Statement}
No Data associated in the manuscript.

\bibliography{sn-bibliography}


\end{document}